\documentstyle[11pt,twoside,cos,epsf]{article} 
 
\heads{Relativistic Astrophysics and Cosmology}{BLACK HOLES FROM PHASE
TRANSITION \ \  {\rm S.G.~Rubin,  
M.Yu.~Khlopov, A.S.~Sakharov}}

\def\beq{\begin{equation}} 
\def\eeq{\end{equation}}

\begin{document} 
 
\Arthead{1}{10}

\Title{PRIMORDIAL BLACK HOLES FROM NON--EQUILIBRIUM SECOND ORDER PHASE TRANSITION}
{S.G.~Rubin$^{(1,2)}$,  
M.Yu.~Khlopov$^{(1,2,3)}$, A.S.~Sakharov$^{(4)}$ \footnote{on the leave
from $^{(1,2)}$}}
{$^{(1)}$ Center for CosmoParticle Physics "Cosmion",  4 Miusskaya pl., 125047 
Moscow, Russia \\ 
$^{(2)}$ Moscow Engineering Physics Institute, Kashirskoe shosse 31, 115409 Moscow, 
Russia \\  
$^{(3)}$ Institute for Applied Mathematics, 4 Miusskaya pl., 125047 Moscow, Russia\\
$^{(4)}$ Labor f\"ur H\"ochenergiephysik, ETH-H\"onggerberg, 
HPK--Geb\"aude, CH--8093 Zurich}   

\Abstract{ The collapse of sufficiently large closed domain wall produced
during second order phase transition in the vacuum state of a scalar field
can lead to the formation of black hole.  The origin of domain walls with
appropriate size and energy density could be a result of evolution of an
effectively massless scalar field at the inflational epoch. We demonstrate
that in this case the situation is valued when there are compact domains
of less favorable vacuum surrounded by a sea of another vacuum. Each
domain has a surface composed of vacuum wall that stores a significant
amount of energy, and can collapse into the black hole. This offers the
way of massive primordial black holes formation in the early
Universe.}

\section{Introduction}
It is well known that any object put within its gravitational radius
forms a black hole (BH). At present time BHs can 
be naturally created only in the result of gravitational collapse of stars
with the mass exceeding three Solar masses \cite{1} in the end of their
evolution. From the other hand, it has long been known
\cite{kp,3,4,mi,ck,klopmop} that primordial black hole (PBH) formation is
possible in the early Universe. PBH can form when the density fluctuations
become larger then unity on a scale intermediate between Jeans length and
horizon size. Other possibilities are related to the dynamics of various
topological defects such as the collapse of cosmic strings \cite{3} from
the thermal second order phase transition or to the collisions of the
bubble walls  \cite{4,mi} created at  the first order phase transitions. 
Formally, there is no limit on the mass of PBH that forms 
after the collapse of highly overdense region, it is only 
needed to form appropriate spectrum of initial density fluctuations at
the inflation~\cite{lk}. We cannot expect {\it {a priori}} the same
"no--mass--limit" condition in the case of PBHs formed by topological 
defects, because the mass of such PBHs is defined by the correlation
length of the respective phase transition. 

In this paper we concern with
the possibility to form PBHs after the self--collapse of closed domain
walls created during a second order phase transition. Such PBHs should
have small masses in the case of thermal second order phase transition
with usual equilibrium initial conditions. It takes place because the
characteristic size of walls coincides with the correlation length of
phase transition, which initiate the formation of that domain walls.
Usually we deal with high temperature phase transition in the early 
Universe, that makes the respective correlation length quite 
small and does not allow to store a significant amount of 
energy within the individual closed vacuum wall. Moreover the phase 
transition can have quite complicated dynamics, namely it can be
accompanied by another phase transition that creates 
another type of topological defects, namely, vacuum
strings~\cite{kim,sik} (see discussion in Section~2). As a result we
get the mesh of vacuum strings connected by walls where the probability
of the existence of closed walls is strongly suppressed~\cite{sik}.   
We discuss here the non--equilibrium scenario of closed vacuum 
walls formation that opens a new possibility for the massive 
PBHs production in the early Universe.

The starting point is rather general and evident. If a potential 
of a system possesses at least two different vacuum states 
there are two possibilities to populate that states in the 
early Universe. The first one is
that the Universe contains both states populated with equal  
probability, that takes place under the usual circumstances of 
thermal phase transition. The other possibility corresponds 
to the case when two vacuum states are populated with 
different probability and there are islands of the less probable vacuum,
surrounded by the sea of another, more preferable, vacuum. The last
possibility can take place when we go beyond the equilibrium conditions,
established by pure thermal dynamics. More definitely, it is necessary
to redefine effectively the correlation length of the scalar field
that drives a phase transition and consequently the formation of
topological defects. We will show that  the only 
necessary ingredient for it is the existence of an effectively 
flat direction(s), along which the scalar potential vanishes 
during inflation. Then the background de-- Sitter fluctuations 
of such effectively massless scalar field could provide non-- 
equilibrium redefinition of correlation length and  give rise 
to the islands of one vacuum in the sea of another one. In 
spite of such redefinition the
phase transition itself takes place deeply in the Friedman-- 
Robertson-Walker (FRW) epoch.
After the phase transition two vacua are separated by a wall, 
and such a wall can be very big. The motion of closed vacuum walls 
has been first driven analytically in~\cite{tk}. At some moment 
after crossing horizon they start shrinking due to surface 
tension. As a result, if the wall does not release the 
significant fraction of its energy in the form of outward 
scalar waves, almost the whole energy of such closed wall can be 
concentrated in a small volume within its gravitational radius what is
the necessary condition for PBH formation.

The mass spectrum of the PBHs which can be created by such a 
way depends on the scalar field potential which parametrizes 
the flat direction during inflation and triggers the phase 
transition at the FRW stage. Through the paper we will deal with so called pseudo 
Nambu--Goldstone (PNG) potential, that is quite common for the 
particle physics models.

The plan of this paper is as following. In Section~2 the 
origin of vacuum walls at the PNG potential is discussed. The 
possibility to form the non--equilibrium conditions for the 
second order phase transition at the inflantional stage is 
considered in Section~3. In Section~4 the minimal conditions 
of PBHs formation from collapsing closed vacuum walls are 
discussed. The mass spectrum of PBHs is evaluated numericaly 
for the certain choises of parameters. 

\section{Domain Walls at the PNG Potential Under Thermal Approach}
Domain walls are planar defects in the vacuum alignment over 
space. They can appear if the manifold of degenerate (or 
nearly degenerate) vacua of the theory is disconnected. This 
is generally the case if there is a discrete symmetry in the 
potential. Such type of symmetry can be fundamental as, for example, in
the double well potential or can arise dynamically during the breakdown
induced by instanton effects of a continuous symmetry (see for example
\cite{kim}). Moreover, discrete symmetry can be got as a result of
explicit breaking of a continuous symmetry due to Yukawa--type couplings
as in the case of classical PNG boson or due to higher--order 
nonrenormalizable interactions that often takes plase in 
string--inspired models \cite{string}. In all this cases the 
resulting term of potential has the form  
\beq 
\label{potential}
V=\Lambda^4\left( 1-\cos\frac{\phi}{f}\right)
\eeq 
where $f$ is the scale of $U(1)$ symmetry breaking. The
potential describing a spontaneously broken $U(1)$ symmetry has 
standard Mexican hat form with bottom radius $f$ and reads
\beq
\label{mex}
\label{3phase}V(\varphi )=\lambda (\left| \varphi \right| 
^{2}-f^{2}/2)^{2},
\eeq
where $\varphi =\frac{f}{\sqrt{2}}\exp {\left( i\phi 
/f\right)}$.
The term (\ref{potential}) breaks the
continuous symmetry down to the symmetry $\phi\to\phi +2\pi nf$, where $n$ is
integer. Thus potential (\ref{potential}) 
has a number of discrete degenerate minima. That minima also 
can be non--equivalent if the manifold spanned by $\phi$ is 
non--compact. Moreover, it is not even necessary for the non--equivalent
minima to be exactly degenerate, so the symmetries we 
consider may be only approximate ones, although non--degeneracy may
change the details of wall evolution \cite{string}. However, as we have
noted, many reasonable models can give rise to cosine potentials like 
(\ref{potential}), therefore we are going to use it in the 
explicit calculations in this paper. 
The equation of motion that corresponds to (\ref{potential}) 
admits kink--like, domain wall solution, which interpolates 
between two adjacent vacua. If we will consider that wall as 
lying in the perpendicular to $x$ axis then the static domain wall
solution between two vacua $\phi_1=0$ and  $\phi_2=2\pi f$ is given
by
\beq
\label{kink} \phi_{wall}(x,x_0)=4f\arctan\left(\exp\frac{x-
x_0}{d}\right)
\eeq
where $x_0$ is the location point of the center of the wall, 
$d$ is its width, 
\beq
\label{size}
d=\frac{f}{\Lambda^2}=m^{-1}
\eeq
and $m$ is the mass of PNG field. The surface energy density 
of the wall reads \cite{vil,sik} 
\beq
\label{dens}
\sigma=\int_{-\infty}^{+\infty}\Lambda^4\left( 1-
\cos\frac{\phi_{wall}}{f}\right)dx\approx 8f\Lambda^2 
\eeq   
The thermal dynamics of a system invoking two 
successive second order phase transitions, with spontaneous 
$U(1)$ symmetry breaking and with the explicit one, results in the
formation of a quite sophisticated 
system of topological defects, network of walls bounded by 
strings (see for review \cite{sik}). We suppose that 
$f>>\Lambda$. This means that first the spontaneous $U(1)$ 
symmetry breaking takes place in the early Universe at the 
temperature $T\simeq f$. After that moment the Universe is 
filled with global $U(1)$ strings \cite{vs,sik} and the phase 
$\theta=\phi /f$ of complex scalar field $\phi$ obeys the 
standard equation of motion  \beq \label{eqm} \ddot\theta 
+3H\dot\theta +\frac{dV}{d\theta}=0
\eeq
As long as the Hubble expansion is fast enough to provide 
the domination of friction term over the gradient one, the 
explicit symmetry is restored. It takes place until the moment 
when the temperature falls down below the value of $\Lambda$. At that 
period the dynamics of global string network implies that there 
is approximately only one string per Hubble horizon 
\cite{vv,vs,sik}, that defines the average distance between 
strings. The correlation length for the second phase transition 
with explicit $U(1)$ symmetry breaking is just equal to the average
distance between global strings at the moment, when it takes place  and
triggers the formation of domain walls \cite{sik,vs}. Thus the typical
size of domain walls created during last phase transition corresponds
to the size of Hubble horizon at the temperature $T\simeq\Lambda$.
As it has been pointed out in the Introduction, the closed 
domain walls could collapse to the black hole by radiating away their
asphericity due to the surface tension forces~\cite{tk,ipsik}. If we
concern the fate of closed domain walls that could be formed after the
last phase transition, we have to compare the gravitational radius 
$r_g=2E/m_p^2$ of the energy $E$ stored in such a closed shell, with the
minimal radius that can be achieved in the result of such collapse. The
typical size of a closed 
wall is given by the correlation length of the last phase 
transition and reads $L_c=0.3m_p/(\sqrt{g_*}\Lambda^2)$ ($g_*$ 
counts the effectively number degrees of freedom in the plasma 
and according to the standard model can not be much less then $10^2$). 
Combining it with (\ref{dens}) the gravitational radius 
will read $r_g\simeq 10^{-3}\frac{f}{\Lambda^2}$.
From the other hand the numerical simulation of the collapse of 
closed domain wall at the PNG potential \cite{we} shows that 
the minimal radius of shrinking cannot be much smaller than the 
wall's width (\ref{size}). This means that it is impossible to 
concentrate the energy of closed walls under the gravitational 
radius by self--collapse if we consider the domain walls as the 
result of usual thermal phase transition. Moreover the system 
of topological defects contains preferably walls bounded 
by strings and open walls, while the probability of closed 
walls formation is strongly suppressed~\cite{sik}.      

To escape such no--go results of equilibrium dynamics we need 
to enlarge essentially the distance between strings and to 
increase the probability
of closed walls formation. The first problem can be solved by 
the inflational blow up of the typical distance of string's separation.  
As well we show below that the non--equilibrium 
conditions imposed by background inflation allow us to bias the 
preference of the theory to choose one vacuum state over 
another that increase essentially the probability of very large 
closed walls formation.

\section{Phase Transition With Non--Equilibrium Initial 
Conditions}
We consider the Universe that, due to the existence of an 
inflaton, goes through a period of inflation and then settles 
down to the standard FRW geometry. In addition 
we introduce a complex scalar field $\varphi$, not the inflaton,  with a
large radial mass $\sqrt{\lambda}f>H_i$, that has got Mexican hat 
potential (\ref{mex}). It makes sure that $U(1)$ symmetry is 
already broken spontaneously at the beginning of inflation in 
our cosmological horizon. Thereby we deal only with the phase 
of that complex field $\theta =\phi /f$, which paramertrizes 
potential (\ref{potential}). Under this condition we come to 
the conclusion, that the correlation
 length of second order phase
transition with spontaneously broken $U(1)$ symmetry exceeds the present
cosmological horizon, and all global $U(1)$
strings are beyond our horizon \footnote{Also we can consider, 
a complex scalar field with a large initial radial positive 
mass that during inflation undergo a phase transition to a 
spontaneously broken $U(1)$ symmetry \cite{lk,lindebook}. In 
such a case the correlation length can be smaller then our 
cosmological horizon, but significantly larger then in the 
case of usual thermal dynamics}. Furthermore, we assume that   
\beq \label{cond} m<<H_i,  \eeq where $H_i$ is the Hubble 
constant during inflation. This condition implies that during 
inflation the potential energy of field $\phi$ is much smaller than the
cosmological friction term what justifies neglecting 
the potential until the Universe goes deeply into the FRW phase.

The dynamics of such effectively massless scalar field 
$\theta$ existing on de Sitter space breaks into two parts 
(see for review \cite{lindebook}). First, there is a 
classical field $\theta_0$, which satisfies the standard 
classical equation of motion (\ref{eqm}).
During inflation, and long afterward, $H_i$ is very large (by 
assumption) compared to the potential (\ref{potential}). It 
follows that we can drop the gradient term in the equation of 
motion (\ref{eqm}) and resulting equation is solved by 
$\theta_0=\theta_{N_{max}}$, where $\theta_{N_{max}}$ is an 
arbitrary constant. We emphasize that there is no information 
in the theory that can fix the value of $\theta_{N_{max}}$. It 
is completely arbitrary. We make standard assumption that our 
present horizon that has been nucleated at the $N_{max}$ e--
folds before the end of inflationary epoch is embedded in an 
enormous inflation horizon, created by exponential blow up of 
a single causal horizon. It follows that $\theta_{N_{max}}$ 
will be the same over the inter inflationary horizon. We put 
$\theta_{N_{max}}<\pi$ without loss of generality.
Second we must consider the quantum fluctuations of the phase 
$\theta$ at the de Sitter background. It is well known 
\cite{alstar}, that there are quantum fluctuations produced on 
the vacuum state of $\theta$ due to the boundary conditions of 
de Sitter space. These fluctuations are
sometimes referred to as contribution to the "Hawking 
temperature" of de Sitter space but, in fact, there are not 
true thermal effects \cite{lindebook}. It makes the dynamics 
of phase $\theta$ strongly non--equilibrium leading to the 
non--thermal distribution of scales populated with different 
vacuums in the postinflationary Universe. 
Such fluctuations can be described by "quasiclassical" scalar 
field, which contains components with wavelengths ranging from 
the size of particle horizon during inflation $H^{-1}_i$, all 
the way out to the inflation horizon $H^{-1}_ie^{N_{max}}$. 
Thus, when the wavelength of a particular fluctuation becomes
greater than $H^{-1}_i$, 
the average amplitude of this 
fluctuation freezes out at some value due to the large friction term
in the equation of motion, whereas its wavelength
grows exponentially. In the other words such a frozen 
fluctuation is equivalent to classical field, discussed earlier, 
that does not vanish after averaging over macroscopic space 
intervals. The vacuum contains fluctuations of every 
wavelengths and hence inflation leads to the creation of new 
regions containing the classical field of different amplitudes 
with scale greater than $H^{-1}_i$. The average amplitude of 
such fluctuations for massless field generated during each 
time interval $H^{-1}_i$ is \cite{alstar}  \beq \label{step} 
\label{dtheta}\delta \theta =\frac{H_i}{2\pi f} \eeq  In the 
other words the phase $\theta$ makes quantum step 
$\frac{H_i}{2\pi f}$ during every e--fold and in every space 
volume with characteristic size of the order of the $H^{-
1}_i$. The total number of steps during time interval $\Delta 
t$ is given by $N=H_i\Delta t$. Such a motion looks like the 
one--dimensional Brownian motion. The $\theta_{N_{max}}$ plays 
the role of the starting point for this  Brownian motion. 
Thus, the initial domain containing phase $\theta _{N_{max}}$ 
increases its volume in $e^3$ times after one e-- fold by 
definition and hence contains $e^3$ separate, causally 
disconnected domains of size $H^{-1}_i$. Each domain is 
characterized by everage phase value 
\beq \label{fluct}
\theta _{N_{max}-1}=\theta_{N_{max}}\pm \delta \theta. 
\eeq
In half of these domains the phases evolve
toward $\pi$ while in the other domains they move toward zero. 
This process is duplicated in each volume of size $H^{-1}$ during 
next e--fold. Now at any given scale $l=k^{-1}$ the size 
distribution of the phase
value $\theta$ can be described by Gauss' law 
\cite{gaus}.  \beq \label{gaus}
P(\theta _l,l)=\frac 1{\sqrt{2\pi }\sigma _l}\exp 
{\left\{ -\frac{(\theta _{N_{max}}-\theta 
_l)^2}{2\sigma _l^2}\right\} }, \eeq
where the dispersion could be expressed in the 
following manner \begin{equation}
\label{b19}
\sigma _l^2=\frac{H^2}{4\pi 
^2}\int\limits_{k_{min}}^kd\ln{k}= \frac{H^2}{4\pi 
^2}\ln {\frac{l_{max}}l=\frac{H^2}{4\pi 
^2f^2}(N_{max}-N_l)}, 
\end{equation}

As long as condition (\ref{cond}) is satisfied the vacuum is well 
defined by the distribution (\ref{gaus}) of fluctuations around 
chosen constant $\theta _{N_{max}}$. Eventually, long after the end 
of inflational epoch, $H(t)$ decreases so significantly that the 
gradient term in the equation of motion (\ref{eqm}) begins to 
dominate over the friction term, and $\theta$ starts to oscillate around
the one of degenerated minimum. We refer the time $t_c$ moment at which
the condition (\ref{cond}) is not valid anymore as the time moment when
the phase transition is triggered. 
The key point of our consideration is based on the following 
observation: At $t_c$ the field configuration of phase $\theta$ 
within the volume of contemporary horizon is uniform and defined by 
$\theta _{N_{max}}$. There are, however, fluctuations due to the 
quasi--classical random field (\ref{gaus}) inside this volume and 
the spectrum of this fluctuations contains all wavelentghs up to 
biggest cosmological scale. Thus the field $\theta$ at $t_c$ feels 
the tilt of potential and must decide to which of the two vacua 
$\theta_{min}=0$ or $\theta_{min}=2\pi$ it should roll down at the 
beginning of oscillation. We deal with the situation when at the 
beginning of inflation the Universe contains the phase 
$\theta_{N_{max}}<\pi$ and hence the final state of the main part 
of the present particle horizon is $\theta_{min} =0$. On the 
contrary, there will be the islands with $\theta >\pi$, which are 
the results of fluctuations (\ref{fluct}) with positive sign. The 
phases inside that islands will move to the final state 
$\theta_{min} =2\pi$ after the triggering of the phase transition.

Thus, the strongly non--equilibrium distribution of the phase,  
causing by inflation, leads to the formation of islands with vacuum 
$2\pi$ in the space with zero phase at the time moment $t_c$. Both 
states are separated by closed walls at $\theta=\pi$. The distribution 
on size for such closed domain walls resembles the size 
distribution of fluctuations (\ref{fluct}) that  consists on the 
steps with positive sign and  move the phase to the region 
$\theta>\pi$ during the inflational  stage. The probability to have 
such fluctuations can be significant that provides us with the
possibility to form quite large number of very big closed domain 
walls. 

\section{Results and Discussion}
As we have seen in the preceding section all regions with phase 
$\theta >\pi$ are converted into islands with vacuum 
$\theta_{\min}=2\pi$ surrounding by the closed walls. The size 
distribution of closed walls imprints the size distribution of 
domains filled with phase coming from fluctuations that have crossed the
point $\pi$ during the one dimensional brownian  motion. The physical
size that leaves the horizon during e-- fold  number N ($N\le N_{max}$)
reads \beq \label{ps} l=H_i^{-1}e^N \eeq This scale becomes comparable
to the FRW particle horizon at the moment 
\beq
\label{ph} t_h=H^{-1}e^{2N} 
\eeq 
It is clear that we will start to 
observe the selfcollapse of a closed domain wall when its size is 
causally connected. Approximately at the same time the wall is 
acquiring spherical form due to the surface tension. Thus, if the 
amount of energy stored in a such vacuum configuration \beq 
\label{mass} E\approx\sigma t_h^2 \eeq
is large enough, the BH can be formed in the result of its
selfcollapse. More definitely, the gravitation radius of 
configuration should exceed the minimal size up to which it can 
collapse. In our case the collapse of closed domain wall, coming 
from potential (\ref{potential}), changes on repulsion at the size 
comparable with wall's width~\cite{we}. This establish cut off for 
the PBH's mass spectrum at the small masses range. 

To evaluate numerically mass spectrum of PBHs we have to calculate 
the size distribution of domains that contains phases, which are at 
the range $\theta >\pi$. Suppose that at e--fold $N=\ln{(lH_i)}$ 
before the end of inflation the volume $V(\bar{\theta},N)$ has been 
filled with phase value $\bar{\theta}$. Then at the e--fold $N-1$ 
the volume filled with average 
phase $\bar{\theta}$ obeys following iterative expression \cite{anti}  
\begin{equation} \label{iteration}V(\bar{\theta},N-
1)=e^{3}V(\bar{\theta},N)+(V_{U}(N)-
e^{3}V(\bar{\theta},N))P(\bar{\theta},N-1)\delta\theta,  
\end{equation} here the $V_{U}(N)\approx e^{3N}H^{-3}_i$ is the 
volume of the
Universe at $N$ e--fold. We applied here distribution (\ref{gaus}). 
Now one can easily calculate the size distribution of domains 
filled with appropriate value of phase corresponding to $N$, with the
use of expression~(\ref{iteration}). For our numerical calculations we
have chosen the following  reasonable values for inflational Hubble
constant $H=10^{13}$~GeV  and for the radius of PNG potential
$f=10^{14}$~GeV (see for  example \cite{frees}). Also we suppose that
$N_{max}=60$ and that the total energy of the wall is smaller than 
the total energy of the medium inside it. 
The simulation has been performed for two cases 
that depend on the $\Lambda$ and $\theta_{N_{max}}$. The results 
are following.

\vspace{0.3cm}
{\centering \begin{tabular}{c|c|c|c|c|c|c|c|c|c}
\hline 
\( \log_{10} \frac{M_{PBH}}{1g.} \)&
12&
13&
14&
15&
16&
17&
20&
21&
22\\
\hline 
\hline 
\( \log_{10} n_{PBH} \)&
20.4&
18.4&
16.3&
14.2&
12.0&
9.79&
5.18&
2.77&
0.30\\
\hline 
\multicolumn{10}{c}{ }\\
\multicolumn{10}{c}{ \( \Lambda =10^{8} \)GeV (see for

example\cite{frees}) , \( \theta _{N_{max}}=0.65 \)}\\ 
\multicolumn{10}{c}{}\\ \multicolumn{10}{c}{Table 1. The 
total number \( n_{PBH} \) of PBHs against their masses \( M_{PBH} 
\).}\\ \end{tabular}\par} \vspace{0.3cm}
\noindent The results of Table 1. are fitted in such a manner 
to satisfy the most stronger astrophysical constraints. It is 
known \cite{21} that only PBHs with masses larger then 
$\approx 10^{15}$~g. can survive in respect to Hawking 
evaporation. The observations of diffused gamma ray background 
establish strong limit~\cite{mg} on the density fraction 
$\Omega_{PBH}<10^{-9}$ of PBHs with masses $10^{14}\div 
10^{15}$~g.. The next limit, which has to be checked is the 
limit on the abundance of PBHs with masses $10^{12}\div 
10^{13}$~g.. Although such light PBHs disappeared already due 
to Hawking radiation it could produce a large amount of 
entropy at the epoch of nucleosinthesis \cite{nas}.
Actually 
the constraints \cite{nas} is valued for the PBH's mass range 
$10^9\div 10^{13}$~g. and reads $\beta <10^{-
15}(10^9g./M_{PBH})$, here $\beta$ is the density fraction of 
PBHs at the moment of their formation. 
It means that at the moment of full evaporation of such PBHs 
$\tau =M^3_{PBH}/(g_*m^4_p)$ their contribution into the total 
density of the Universe should satisfy the following 
restriction $\alpha <10^{-4}$.  To check our spectrum we 
express the density fraction of PBHs at the moment 
of their full evaporation in the terms of $n_{PBH}$, it gives 
$\alpha\simeq 10^{-8}n_{PBH}(1g./M_{PBH})^2$. Thus in the  case of Table 1. the entropy
production limit~\cite{nas} is  also satisfied. The more careful
consideration of particle content of $10^{10}\div 10^{13}$g. PBH decay
products shows that such PBHs are
very effective sources of antiprotons at the period of their
evaporation~\cite{ck}. According to the theory on non--equilibrium
nucleosynthesis (see~\cite{klopmop} and references therein) it can change
dramatically the abundance of such elements as $^3He$, $^6Li$ and $^7Li$.
This fact establishes the most stringent constraints on the density
fraction of PBHs in the mass rage $10^{10}\div
10^{13}$g.~\cite{klopmop}~\footnote{Here we take into account only the
standard model set of particle spices. The invoking of the some
supersymmetric extensions of the standard model with a large number of
unstable moduli fields which could come from evaporating PBHs also would
influence on the primordial chemical content of the Universe (see for
example~\cite{moduli}).}. Thus, taking into account the effects of non--equilibrium
nucleosynthesis of $^3He$ and lithium we have got following
limits on the density fraction of PBHs with mass range indicated
above: $\alpha_{^3He}<10^{-15}\Omega_b(M_{PBH}/1g.)k^{-1}$, 
$\alpha_{Li}<10^{-19}\Omega_b(M_{PBH}/1g.)^{5/4}k^{-1}$, where the
$\Omega_b$ is the fraction of baryon density and $k$ varies from $1/4$ to
$1/6$ for PBHs of different masses~\cite{klopmop}. These limits are also
satisfied our model in Table 1. As well we can
see  from the Table 1. that the number of BH with masses more than
$10^{22}$~g. is negligible, whereas BHs with masses smaller than
$10^{12}$~g. were not produced at all because of their small gravitational
radius. 

Another interesting case is small $\Lambda$ limit that comes from
QCD~\cite{kim}. It gives rise  to massive BH.

\vspace{0.3cm}
{\centering \begin{tabular}{c|c|c|c|c|c|c|c|c}
\hline 
\( \log_{10} \frac{M_{PBH}}{1g.} \)&
28.3&
29.5&
30.6&
31.8&
33.0&
34.1&
35.3&
36.4\\
\hline 
\hline 
\( \log_{10} n_{PBH} \)&
17.4&
13.0&
11.1&
9.2&
7.2&
5.1&
3.0&
0.66\\
\hline 
\multicolumn{9}{c}{ }\\
\multicolumn{9}{c}{ \( \Lambda =1 \)GeV (see for exapmle~\cite{kim}) , \(
\theta _{N_{max}}=0.8 \)}\\ \multicolumn{9}{c}{}\\
\multicolumn{9}{c}{Table 2. The total number \( n_{PBH} \) of PBHs against
their masses \( M_{PBH} \).}\\ \end{tabular}\par}
\vspace{0.3cm}

\noindent In this case the wall width is very
large and so that only rather massive BHs could be formed. This fact can
be applied for explanation of giant BHs origin in the centres of galaxies
(see for example~\cite{agn})  The maximal PBH masses in the Table 2 are at
least by 3--4 orders of the magnitude smaller than the masses of BHs,
assumed to be present in the centres of galaxies.
However the account for possible strong concentration of small mass PBHs
around close-to-maximal mass PBHs can provide the evolution and collapse
of PBH systems into black holes with the mass, exceeding by several orders
of the magnitude the maximal PBH mass. It provides the possibility
to evolve the approach to the origin of AGNs on the base of our model. 

To complete our discussion let us mention following. We assumed that
the coherent  field oscillations around the true vacua, which start after
the triggering of phase transition with the amplitude
proportional to the residual between the local value of phase and the
local vacuum phase, are damped due to relatively strong dissipation, so
that no energy density is stored in the form of such oscillations. If the
dissipation is not effective, as it is the case for invisible
axion~\cite{kim}, the inhomogeneity of the phase distribution before the
phase transition results in the inhomogeneity of the energy density
distribution of coherent field  oscillations. Then such energy density
contributes into the total cosmological density and its large scale
inhomogeneity induces effects of anisotropy of relic radiation. In this
case one can strongly constrain the model parameters from the
observational upper limits on the total  cosmological density and on the
possible anisotropy of thermal  background radiation induced by
isocurvature perturbations~\cite{iso}. 

Leaving a more completely study of the last two issues to future
publications, we can conclude that the non--equilibrium
dynamics of scalar field with effectively flat direction can give a new
spin to the PBHs formation mechanisms applying collapse of 
domain walls.
\\
\\
{\it Acknowledgements}. The work of SGR and MYuK was
partially performed in the framework of Section "Cosmoparticle 
physics" of Russian State Scientific Technological Program 
"Astronomy. Fundamental Space Research", with the support of Cosmion-ETHZ
and Epcos-AMS collaborations.  ASS and MYuK acknowledge supporte from
Khalatnikov--Starobinsky school (grant 00--15--96699).

\end{document}